\documentclass[preprint]{aastex}

\usepackage{amsmath}
\usepackage{amssymb}
\usepackage{epstopdf}
\usepackage{graphicx}
\usepackage{float}
\usepackage{natbib}
\usepackage{footnote}

\newcommand\ora[1]{\overrightarrow{#1}}
\newcommand\abs[1]{\left|{#1}\right|}

\slugcomment{Submitted to PASP}

\shorttitle{Multiplexed Imaging}
\shortauthors{Zackay and Gal-Yam}

\begin{document}


\title{The Multiplexed Imaging Method \\
    High-Resolution Wide Field Imaging Using Physically Small Detectors}


\author{Barak Zackay\altaffilmark{} and Avishay Gal-Yam\altaffilmark{}}
\affil{Weizmann Institute of Science}

\email{bzackay@gmail.com}
\email{avishay.gal-yam@weizmann.ac.il}




\begin{abstract}
We present the method of multiplexed imaging designed for astronomical observations of large sky areas in the IR, visible and UV frequencies. Our method relies on the sparse nature of astronomical observations. The method consists of an optical system that directs light from different locations on the focal plane of a telescope onto the same detector area and an algorithm that reconstructs the original wide-field image. In this way we can use a physically small detector to cover a wide field of view.
We test our reconstruction algorithm using public space telescope data.
Our tests demonstrate the reliability and power of the multiplexed imaging method.
Using our method it will be possible to increase the sky area covered with space telescopes by 1-3 orders of magnitude, depending on the specific scientific goal and optical parameters. 
This method can significantly increase the volume of astronomical surveys, including search programs for exoplanets and transients using space and ground instruments.

\end{abstract}



\section{Introduction}\label{sec:Introduction}

\subsection{Observing a large area of the sky}
Surveying a large sky area is one of the most common and elementary types of observation. In principle, one would
 want to image as wide an area of the sky as possible, at high spatial resolution and through a telescope with a large aperture.
Covering a wide field at high resolution requires a detector with a large physical area and many pixels, leading to high cost and complexity.

Astronomers deal with this problem in two ways:
\begin{itemize}
\item Using complex and expensive arrays of detectors.
\item Using detectors with large pixels, at the expense of resolution.
\end{itemize}
We propose a novel method to address this issue.

\subsection{The origin of noise in astronomical observations}
The performance of the multiplexed imaging method strongly depends on the properties of the relevant noise. We consider the following origins of noise in astronomical observations.
\begin{itemize}
\item \textbf{Poisson noise} - The difference between the expected flux of energy coming from a light source (or a noise source) and the flux actually arrived.
\item \textbf{Background noise} - The fluctuations in the number of photons originating in the space between the observer and the source (e.g, atmospheric background noise and zodiacal light).
\item \textbf{Read noise and instrumental thermal noise (dark current)}: The noise coming from the process of reading the data from the detector and the electrons that originate in the electronics and the detector itself.
\end{itemize}
When imaging from the ground, all types of noise exist and contribute, while when imaging from space, in some situations we can neglect the background noise and assume that the only sources of noise are the read noise and the Poisson noise.


\subsection{Sparsity of astronomical images}
A very important feature of many astronomical images is that they are almost \textbf{empty}. When we pick a random patch of sky and observe it (not a specifically chosen close galaxy, nebula or dense star cluster) there are very few objects with non-zero flux. Most of them are either point source objects (the size of the seeing disk) or small patches (like distant galaxies) with sizes on the scale of a few arcseconds.
Our analysis only applies to sparse images because the source of the improvement that we offer is the sparsity of the part of the sky observed.

\section{The Multiplexed Imaging Method}\label{sec:method}
\subsection{The general idea}
We propose to use the sparse nature of astronomical images to effectively measure all objects contained in the corrected field of view of a telescope using a physically small detector, without reducing the resolution. This is done by projecting different areas of the focal plane simultaneously onto the detector. Because of the scarcity of objects one can perform scientific measurements using the combined images with the same quality and greater efficiency compared to mosaicking. 
Flux measurements of known sources in combined images can be done directly (e.g to search for transients ans planets). Using sub-observations it is possible to chart an unknown part of the sky. 

In section \ref{sec:Recovery} we analyze both cases and present our recovery algorithm. In section \ref{sec:Efficiency} we analyze the efficiency gain factor of our method under different dominant noise sources and show that multiplexed imaging is very efficient when the dominant noise source is either Poisson or read noise.
In section \ref{sec:Simulations} we show simulations of our charting recovery algorithm.
In section \ref{sec:Applications} we analyze the possible impacts of the method on instruments designed for sky surveys, finding exoplanets, looking for transients, fast photometry and lucky imaging, finding orders of magnitude potential improvements to the area observed per unit time.
A conceptual optical realization of this method is discussed in Ben-Ami et al. (in preparation), and laboratory experiments with this design are ongoing. 
\subsection{Definitions}
\begin{figure}[H]
\centering
\includegraphics[width=50mm]{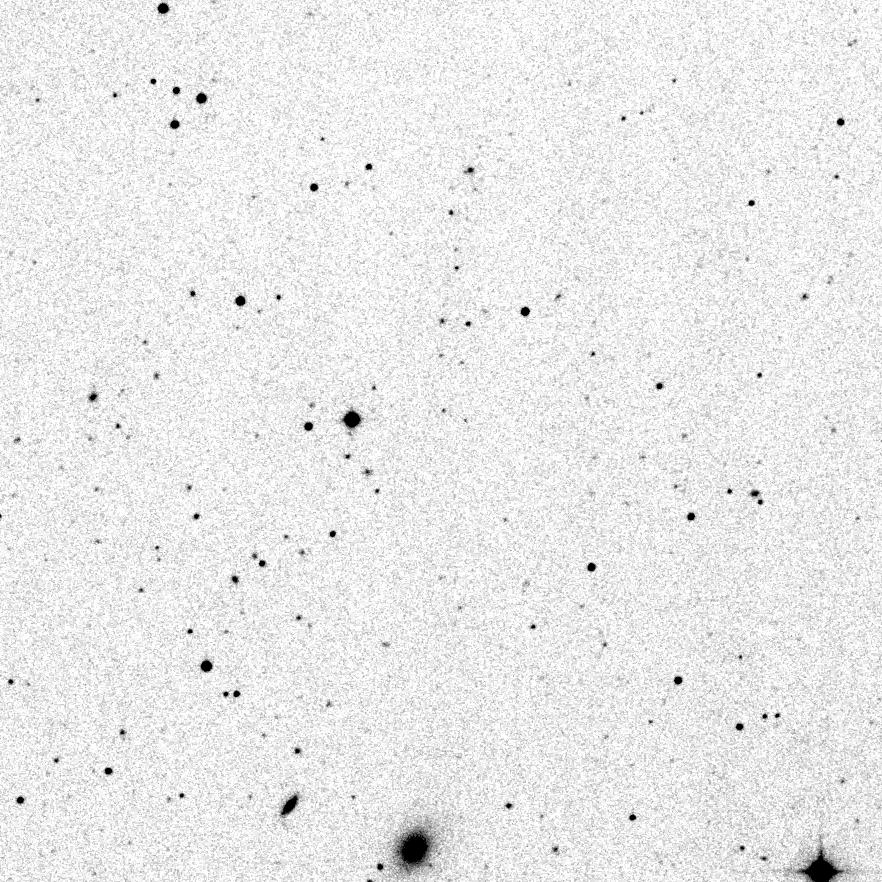}
\includegraphics[width=50mm]{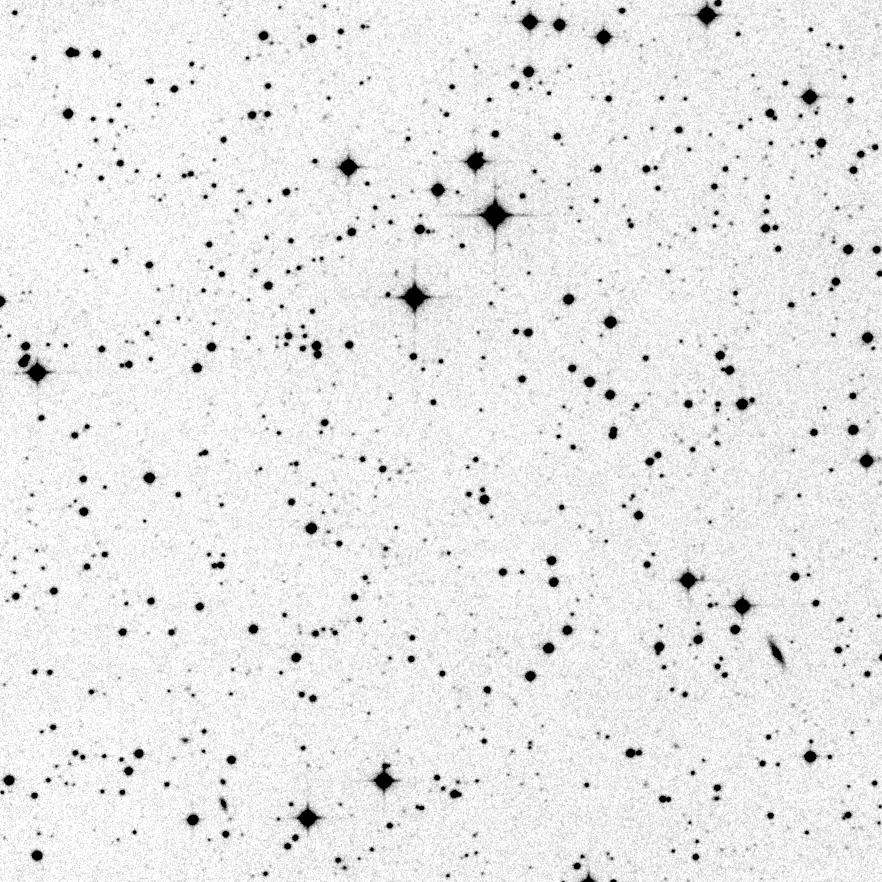}
\includegraphics[width=50mm]{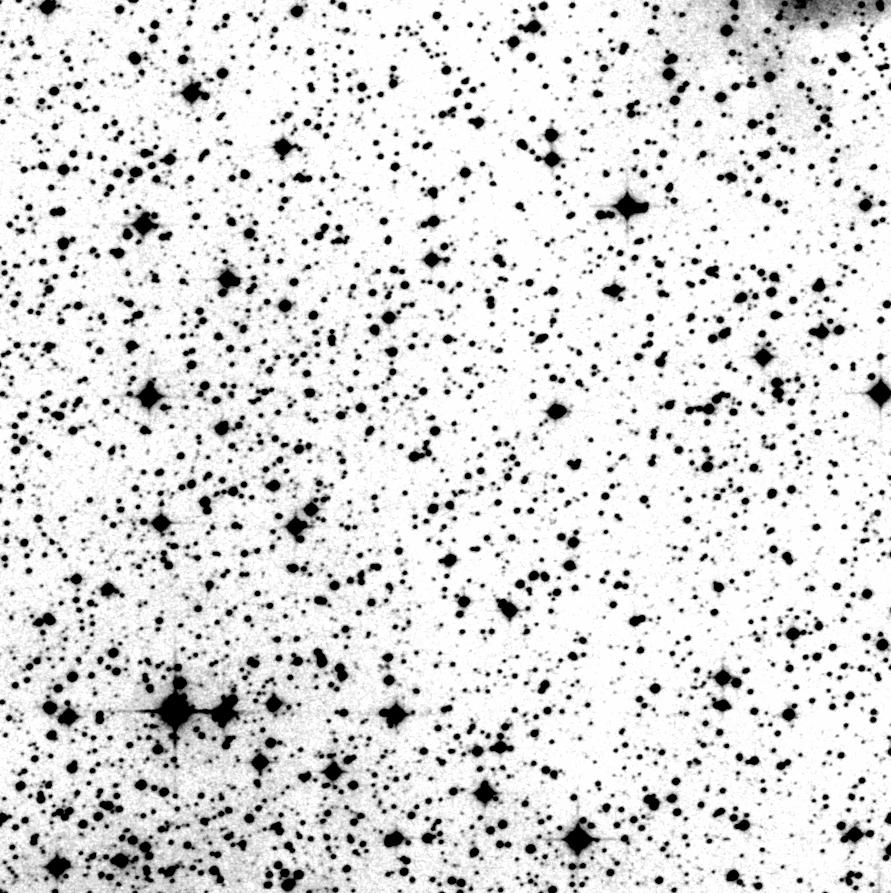}
\caption{The sparsity of astronomical images. \newline DSS\protect\footnote{} images representing the sparsity variation of astronomical images across the sky. Left: the galactic pole with $Pd \approx \frac{1}{150}$. Middle: a typical region at galactic latitude $19.1^\circ$ with $Pd \approx \frac{1}{34}$. Right: the galactic center with $Pd \approx \frac{1}{7}$.
It is important to note that the density estimate $Pd$ depends on the depth of the image, the resolution and the seeing, and therefore imaging the same area with different instruments might yield different densities.\label{fig:Emptiness}}
\end{figure}

\begin{itemize}
  \item The focal plane is divided into $N$ parts, each part can be covered by the detector in a single exposure of length $T$.
  \item Our device directs the light of $K$ parts ($K<N$) onto one detector.
  \item An observation is composed of $M$ sub-observations. Each sub-observation is a measurement of the sum of the flux from $K$ areas on the focal plane.
  \item The time each observation takes depends on the exposure time $T_e$, readout time $T_r$ and slew time $T_s$. When referring to two different exposure times, the duration of multiplexed imaging will be denoted by $T^*_e$. 
  \item The total time required for one observation using multiplexed imaging is given by $$T^*_{total} = M(T^*_e+T_r)+T_s$$
  \item Regular imaging can be thought of as multiplexed imaging with $K=1$ and $M=N$ (it will take $N$ observations to cover the whole area). In that case $T_{total}=N(T_e+T_r+T_s)$.
  \item The efficiency $E$ of the system is given by the time required to cover the described area with the regular mode divided by the time required to do so with the multiplexed method, \textbf{when both observations have the same SNR}, meaning $$E = \frac{N(T_e+T_r+T_s)}{M(T^*_e+T_r)+T_s}$$
  (and $T_e^*$ is adjusted to match the SNR).
  \item For example, under the assumptions that \begin{equation}\label{condition:TeTr} 
  T_e>>T_s+T_r 
  \end{equation} (which is not always the case, see section \ref{subsec:background}) and $MT^*_e = T_e$ (which will make the SNR equal when the dominant source of noise is Poisson noise)  this yields $E = \frac{N(T_e+T_r+T_s)}{M(T^*_e+T_r)+T_s} \approx \frac{NT_e}{T_e}\approx N$. 
  \item We will define the object surface density $d$ to be the number of sources in one part of the sky divided by the area of sky observed.
  \item Denote by $P$ the average (over sources with different intensities and sizes) number of pixels that have a statistically significant contribution per object.
  \item The assumption that a sky area is sparse means that $Pd<<1$ (Fig. \ref{fig:Emptiness}).
  \item The best possible multiplexing and the absolute upper limit on $K$ satisfies $KPd \approx 1$, which means that we are measuring a non-trivial flux with every pixel of our detector.
  \item Generally $Pd$ depends on many parameters such as the depth of the observation, the filter, the field observed, the plate scale and the seeing. $Pd$ values for a few common surveys are:
  \begin{itemize}
  \item 80s Near-UV exposures with GALEX \citep{GALEX} toward high galactic latitudes (see section \ref{sec:Simulations}) have $Pd\sim \frac{1}{1000}$. 
  \item For SDSS \citep{SDSS} g-band imaging toward the north galactic pole we measure $Pd\sim \frac{1}{500}$. 
  \item For PTF \citep{PTF2} single 60s r-band exposures toward the galactic pole, $Pd \sim \frac{1}{100}$.
  \end{itemize} 
\end{itemize}

\footnotetext{http://archive.stsci.edu/cgi-bin/dss\_form}

\begin{figure}[H]
	\plottwo{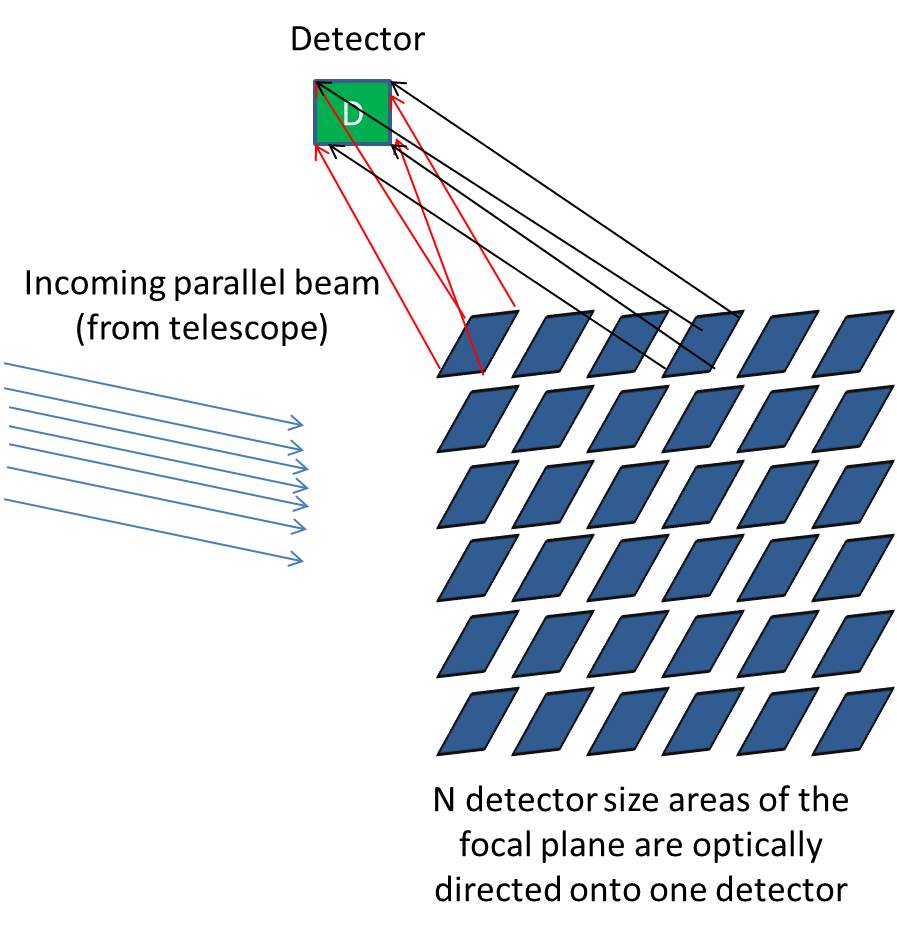}{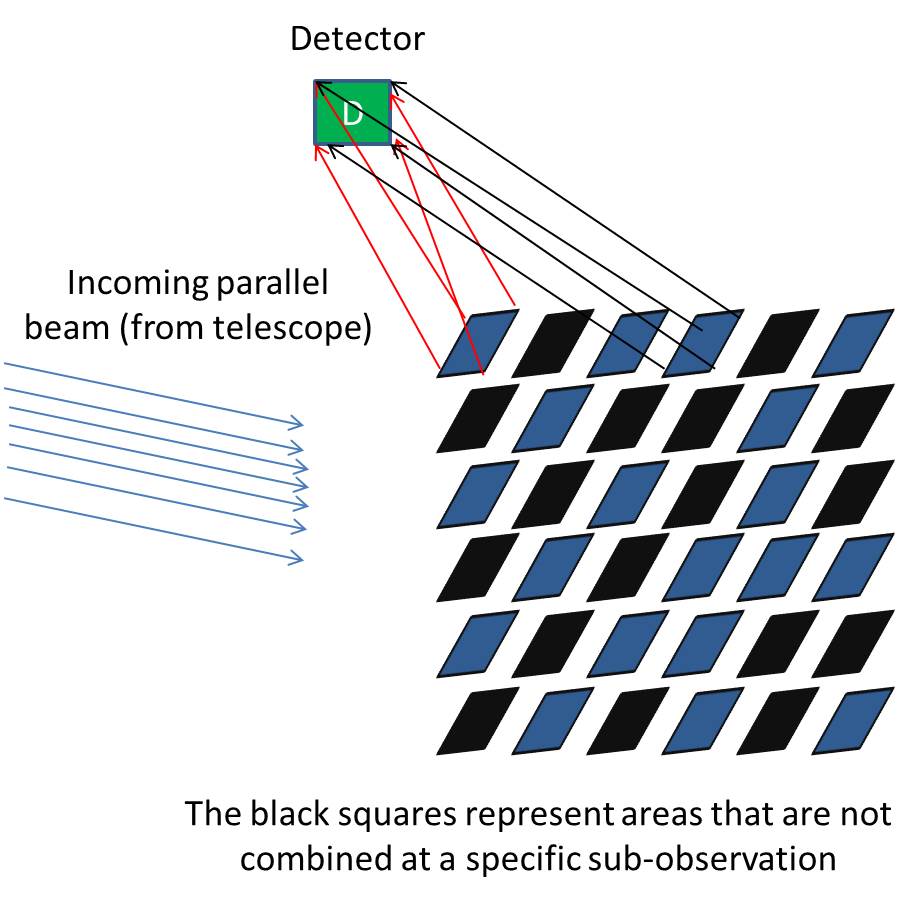}
	\caption{Setup Illustration: $N$ Areas on the focal plane are optically combined onto one detector. In one sub-observation only a subset of $K$ focal plane areas (shown on the right) is directed to the detector.}
	\label{Illustration}
\end{figure}

\section{Recovering the original observation}\label{sec:Recovery}
We consider separately two scientific cases: \textbf{charting} and \textbf{re-observing}.

\subsection{Charting}
We define charting as an observation of a part of the sky that is unknown to the resolution and depth in question.
In this mode, since we have no prior knowledge of the position of every object, we need to obtain several sub-observations, allowing for each part of the sky to have a specific pattern of appearance (Fig. \ref{ToyExample}) in order to recover the correct position of each source.

\subsection{Re-observing}
When re-observing, we have a prior image of the relevant region of the sky. Our observational goal in re-observing mode is to measure the flux from previously known objects, measuring variability or searching for new transients.
In the re-observing mode we can calculate for each pixel on the image a-priory (using our known mapping of the sky) which areas on the focal plane contribute to the measured flux, allowing for a simple recovery algorithm.
The number of sub-observations required and therefore also the efficiency depends on whether the scientific mission requires to measure \textbf{all} the objects in the field, or just as many of them as possible.
Those two cases are analyzed separately. The case where it suffices to measure only most of the objects should allow using a trivial recovery algorithm, and requires only $\approx 1$ sub-observations.
The other case, where all objects in the field are important, is similar to the charting mode because different objects falling on the same detector area must be dealt with. This case is not addressed in this paper as we expect it to be rare.

With recent advances in the recording of a multi-wavelength static image of much of the sky, e.g., by surveys such as SDSS \citep{SDSS}
PS1 \citep{PS1} and PTF (\citealp{PTF1}, \citealp{PTF2}) in the optical, GALEX 
\citep{GALEX} 
in the UV and 2MASS \citep{2MASS}, UKIDSS \citep{UKIDSS} and WISE \citep{WISE} in the IR, the re-observing mode is likely to be the common mode (see applications \ref{subsec:Lucky imaging}, \ref{subsec:Fast photometry}, \ref{subsec:Searching for transients}, \ref{subsec:Planet search}).

\subsection{Construction of the sets}

The recovery algorithm is simple, but to present it, we first need to introduce some notations and construct the sets of regions of sky combined during each sub-observation.
We use $N$, $M$, and $K$ as defined before.

Denote the set of sky regions combined during sub-observation $0 \leq i<M$ by $C_i$. Denote the flux recorded in sub-observation $i$ at pixel location $x$ by $f_i[x]$ and denote the flux arriving to pixel $x$ from region $j$ on the focal plane by $g_j[x]$.
The expected flux (without noise) at each pixel is therefore $$f_{i}[x]= \sum_{j\in C_i}{g_j[x]}$$
Denote for each region $j$ on the focal plane, the representing vector (a binary vector of length $M$) $v_j\in \left\{0,1\right\}^M$ indicating if the flux from region $j$ is combined during sub-observation $i$.
$$v_j[i] = \left\{ \begin{array}{cc}
0&j\notin C_i \\
1&j\in C_i
\end{array} \right. $$
And denote the set of representing vectors $V = \{\overrightarrow{v_i}\}^{N-1}_{i=0}$.
The representing vectors determine uniquely the set of regions that are included in each sub-observation and they can be chosen by the algorithm designer ahead of making the observation. This allows us to choose in a special way the set of vectors such that there will be no ambiguity in the reconstruction algorithm. 

Denote the vector of measured fluxes from all sub-observations at a pixel $x$ on the detector by $$\overrightarrow{\mu} = (f_0[x],f_1[x],\dots,f_{M-1}[x])$$
If a specific pixel $x$ has exactly one non-zero flux contribution coming from a specific sky region $j$, then there exists a real number $a$ such that $$\overrightarrow{\mu} = a\overrightarrow{v_j}$$

\subsubsection{A simple example}
\begin{figure}[ht]
  \centering
  \includegraphics[scale=0.5]{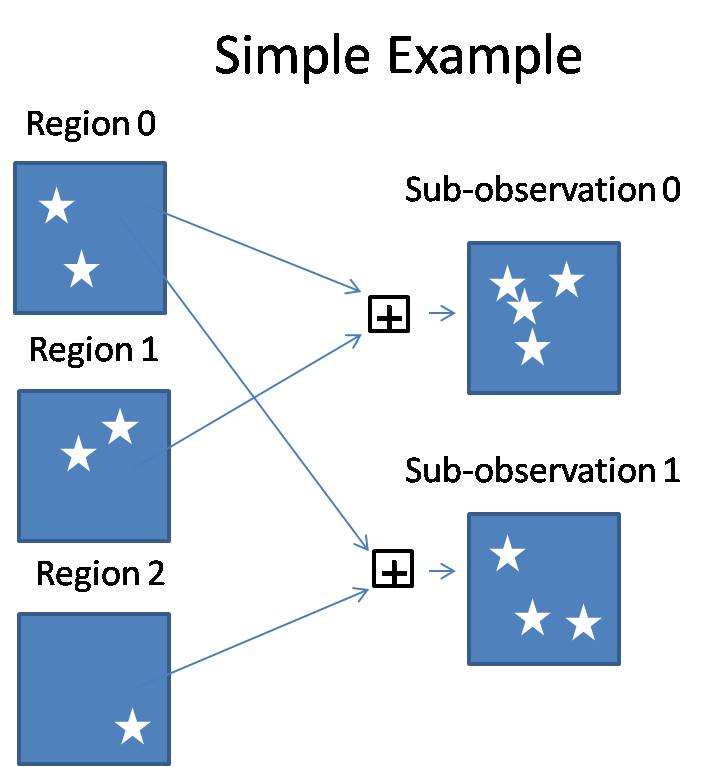} 
  \caption{A simple example showing a multiplexing scheme with $N=3$, $K=2$, $M=2$. }\label{ToyExample}
\end{figure}

When constructing the sets, the recovery ambiguity problem has to be dealt with, and to demonstrate it we will use a simple set of parameters:
we choose $N=3$, $K=2$ and $M=2$. 
We will denote the focal plane sub-areas by $\{0,1,2\}$ as illustrated in Fig.\ref{ToyExample}.

The sets we use are $C_0 = \{0,1\}$ and $C_1 = \{0,2\}$.
The sub-observations we use are therefore $f_0[x] = g_0[x]+g_1[x]$ and $f_1[x] = g_0[x]+g_2[x]$.
The representing vectors will be $\ora{v_0} = (1,1)$, $\ora{v_1} = (1,0)$, $\ora{v_2} = (0,1)$.
For every pixel location $x$ we can construct $\ora{\mu} = (f_0[x],f_1[x])$.
\newline If for a pixel location we observe $$\ora{\mu} = (a,0) = a(1,0) =a\ora{v_1}$$ we can deduce that $g_1[x] = a$ and $g_0[x]=g_2[x]=0$.
\newline If for a pixel location we observe $$\ora{\mu} = (0,a) = a(0,1)=a\ora{v_2}$$ we can deduce that $g_2[x] = a$ and $g_0[x]=g_1[x]=0$.
\newline If for a pixel location we observe $$\ora{\mu} = (a_1,a_1) = a_2(1,0) + a_2(0,1) + (a_1-a_2)(1,1) = a_2\ora{v_1} + a_2\ora{v_2} + (a_1-a_2)\ora{v_0}$$ we can deduce that $g_0[x] = a_1-a_2$ and $g_1[x]=g_2[x]=a_2$. \newline
This demonstrates a possible ambiguity, even without considering observational errors. There could be more than one combination of fluxes that will generate the same observed vector. This problem is inherent because we have less measurements than free parameters we are trying to measure (here we try to measure 3 free parameters, $g_0[x],g_1[x],g_2[x]$, but we have only 2 measurements, $f_0[x],f_1[x]$).
This is where the sparsity assumption is necessary. We assume the original image is sparse, meaning that most of the measured parameters are $0$, and therefore we assume the correct recovery in this case is $g_0[x] = a_1$ and $g_1[x]=g_2[x]=0$.
\newline If for a pixel location we observe $\ora{\mu} = (a_1,a_2)$ we cannot recover the original fluxes because the following cases are equally likely:
\begin{itemize}
\item $g_0[x] = 0$, $g_1[x]=a_1$ and $g_2[x]=a_2$
\item $g_0[x] = a_1$, $g_1[x]=0$ and $g_2[x]=a_2-a_1$.
\end{itemize}
In this case we cannot solve the ambiguity and we cannot determine neither the locations nor the fluxes of the non-zero sources.
We should therefore construct the sets $C_i$ carefully, preventing ambiguities when few sources are contributing non-zero flux to a pixel location.

\subsubsection{Preventing the ambiguity}
The most basic requirement on our sets is that if there is only one non-zero flux contributing to a location $x$ then the recovery is unique.
From here we can immediately deduce a strict demand on the construction of the set $V$:
For every pair of different vectors, $\overrightarrow{v},\overrightarrow{w}\in V$ and for all pairs of real numbers $a,b\neq 0$ the following should hold: $$a\overrightarrow{v}\neq b\overrightarrow{w}$$
This demand, implying $\overrightarrow{v}\neq \overrightarrow{w}$ and $\overrightarrow{v}\neq \overrightarrow{0},\forall v \in V$, defines the absolute lower bound for the number $M$ of sub-observations to chart $N$ regions of the focal-plane, which is $M\geq log_2(N+1)$.

It is sub-optimal to recover only objects that do not overlap with other objects (this limits the multiplexing number $N$ (and therefore $K$) to be smaller than desired, leading to smaller region of the sky being observed).
Therefore we wish to construct our sets such that unique recovery is guaranteed also when two objects fall on the same detector area, (and with high probability will be unique even when there are 3 or more objects falling on the same detector area).
In the case of two objects, in a similar fashion, we have $$\overrightarrow{\mu} = a_1\overrightarrow{v_{i_1}} + a_2\overrightarrow{v_{i_2}} $$
Again, we choose the set $V$ to hold an analogue condition:
for every quadruplet of different vectors, $\overrightarrow{v_{i_1}},\overrightarrow{v_{i_2}},\overrightarrow{v_{j_1}},\overrightarrow{v_{j_2}}\in V$ and for all quadruplets of real numbers $a_1,a_2,b_1,b_2$ the following should hold: \begin{equation}\label{couplesCondition}
a_1\overrightarrow{v_{i_1}}+a_2\overrightarrow{v_{i_2}}\neq b_1\overrightarrow{v_{j_1}} +b_2\overrightarrow{v_{j_2}} 
\end{equation}
In our simple example above, this condition does not hold, as 
$$
\ora{v_1} + \ora{v_2} = \ora{v_0}
$$
This fact is the cause of the ambiguity in the recovery.

When we consider the charting of weak sources, another source of confusion can be the noise (of all kinds). If two regions of sky have close representing vectors, i.e $||\ora{v_i}-\ora{v_j}||^2$ is small, then it is easy to confuse flux coming from region $i$ with flux coming from region $j$. This is because the recovery algorithm can use only sub-observations that contain only one of the regions $i,j$ to distinguish between sources coming from region $i$ and sources coming from region $j$. The SNR of one sub-observation is lower than the SNR of the reconstructed observation, meaning that statistically significant sources on the reconstructed image might not be significant in one sub-observation. This means that weak (yet statistically significant sources) which are non-significant in a single sub-observation can be mistakenly misplaced to positions with a representing vector which is close to the representing vector of the correct position.
Therefore we prefer to choose the representing vectors such that the difference between every pair of vectors is non-zero in at least $r$ coordinates.
\begin{equation}
\label{distanceCondition}
||\ora{v_i}-\ora{v_j}||^2\geq r
\end{equation}

Constructing a set for which conditions \eqref{couplesCondition} and \eqref{distanceCondition} hold is non-trivial, and is discussed in section \ref{subsec:construction}, we will just remark that it can be done for a multiplexing of $N$ with $M\approx 2log_2(N)$ if we want a robust recovery. In a slightly less robust case, when in the above notation $\frac{|a_1-a_2|}{\sqrt{\sigma_{noise}^2 + a_1}}>\gamma \approx 5$ ($\gamma$ is introduced later and is a confidence parameter of the algorithm, this condition means that with high probability there is no confusion between $a_1$ and $a_2$ due to noise) then we can use as little as $M\approx log_2(N)+log_2(log_2(N))$ sub-observations for full recovery.
On our simple example above, the minimal set that satisfies condition \eqref{couplesCondition} and condition \eqref{distanceCondition} with $r=2$ is:  
$$
V = \{(0,1,1),(1,0,1),(1,1,0)\}
$$

\subsubsection{The charting recovery algorithm}
A detailed description of the charting recovery algorithm can be found in section \ref{Algorithm}, 
here we show only a simplified conceptual version.
The input of this algorithm is $\overrightarrow{\mu}$ for every pixel position $x$. The output is the fluxes, $a_0,a_1,\dots$ and the locations they are coming from, $i_0,i_1,\dots$ contributing to pixel $x$.
\newline
\textbf{Algorithm} (with confidence parameter $\gamma$)
\begin{enumerate}
    \item Go over all vectors $\overrightarrow{v_i}\in V$ and find the optimal $a>0$ and $i$ such that $\|\overrightarrow{\mu} - a\overrightarrow{v_i} \|^2$ is minimal. Remember the best position $i$ and flux $a$. If $a<\gamma\sigma_{noise}$ stop and output all pairs of $i,a$ that were found before.
	 \item Update $\overrightarrow{\mu} = \overrightarrow{\mu} - a\overrightarrow{v_i} $
	 \item Go to step 1.
\end{enumerate}
This algorithm is the simplest recovery algorithm one could think of, yet it captures the essence of the algorithm presented in section \ref{Algorithm}, which is statistically more accurate. 
This simple algorithm does not treat the case $a_1=a_2$. We note that using the information from neighboring pixels we can easily solve ambiguities rising from the case $a_1=a_2$. 
The results of employing this algorithm (without using data from neighboring pixels) to real data from GALEX are shown in section ~\ref{sec:Simulations}.

\section{Efficiency analysis}\label{sec:Efficiency}

To understand the behavior of the efficiency $E = \frac{N(T_e+T_r+T_s)}{M(T^*_e+T_r)+T_s} $, we need to know the exposure time factor $\frac{T^*_e}{T_e}$. This factor is determined by the constraint on $E$ that we compare the time needed for observations with equal SNR.  The exposure time factor changes when different noise sources are dominant and for various observing modes (re-observing vs charting), and therefore each case should be handled separately. The efficiency for all cases (assuming condition \eqref{condition:TeTr}) is summarized on table \ref{EfficiencyTable}.
\begin{table}[H]
\begin{center}
\begin{tabular}{|l|l|l|l|}
  \hline
   \raisebox{-0.5em}[1em][1em]{Observing mode} \Huge $\backslash$ \normalsize \raisebox{0.5em}[1.5em]{Dominant noise}  & \raisebox{0.5em}[1.5em]{Poisson} & \raisebox{0.5em}[1.5em]{Background noise} & \raisebox{0.5em}[1.5em]{Read-noise}   \\
  \hline
  Re-observing & $K$& $1$ (*) & $K$ \\
  \hline
  Charting     & $K$& $1$ (*) & $\frac{K}{\sqrt{log_2(N)}}<E = \sqrt{\frac{NK}{M}}<K$ \\
  \hline
\end{tabular}
\end {center}
\caption{Method efficiency for various modes and noise property combinations.
\newline (*)Note that in section \ref{subsec:background} we show that the method can improve the efficiency of background dominated observations if condition \eqref{condition:TeTr} is invalid and $T_r + T_s$ is not negligible compared to $T_e$.  \label{EfficiencyTable}}
\end{table}

Denote the background noise in observation $i$ coming from region $j$ at pixel $x$ as $b_{i,j}[x]$, denote the read noise in observation $i$ at pixel $x$ by $r_i[x]$. We will use the notation $P(\lambda)$ to denote a Poisson random variable with expectancy $\lambda$ (where $\lambda$ is in units of photons). We will assume that all are independent random variables and that the background and read noise have mean $0$ and standard deviations $\sigma_b,\sigma_r$ respectively (and otherwise subtract the mean).
To calculate the SNR of an observation, assuming that only one region $\bar{c}$ contributes non-zero flux $g_{\bar{c}}$ to pixel $x$, lets first look at a specific sub-observation: $$f_i[x]=P(\sum_{j\in C_i}{g_j[x]T^*_e})+\sum_{j\in C_i}{\sqrt{T^*_e}b_{i,j}[x]}+r_i[x]=P(g_{\bar{c}}T^*_e)+\sum_{j\in C_i}{\sqrt{T^*_e}b_{i,j}[x]}+r_i[x]$$
For each region $\bar{c}$ we have $\left|\{i\;s.t\; \bar{c}\in C_i\}\right|\approx \frac{MK}{N}$ (the number of sets $C_i$ that contain $\bar{c}$) sub-observations containing it (in each sub-observation we observe $K$ parts, there are $M$ such sub-observations and the total number of parts is $N$).
Assuming the common case that only the region $\bar{c}$ contributes non-zero flux to pixel $x$, the best SNR of region $\bar{c}$ is achieved when taking the average of all sub-observations $i$ for which $\bar{c}\in C_i$, meaning that the best flux estimation of region $\bar{c}$ is
$$ F_{\bar{c}} = \frac{\sum_{\{i\;s.t\; \bar{c}\in C_i\}}{P(T^*_e g_{\bar{c}})} + \sum_{\{i\;s.t\; \bar{c}\in C_i\}}{\sum_{j\in C_i}{\sqrt{T^*_e}b_{i,j}[x]}} + \sum_{\{i\;s.t\; \bar{c}\in C_i\}}{r_i[x]}}{\left|\{i\;s.t\; \bar{c}\in C_i\}\right|} $$
Multiplying the signal by a factor does not change the SNR so
\begin{equation}\label{eq:SNR}
 \frac{MK}{N}F_{\bar{c}} \approx |\{i\;s.t\; \bar{c}\in C_i\}|F_{\bar{c}} = \sum_{\{i\;s.t\; \bar{c}\in C_i\}}{P(T^*_e g_{\bar{c}})} + \sum_{\{i\;s.t\; \bar{c}\in C_i\}}{\sum_{j\in C_i}{\sqrt{T^*_e}b_{i,j}[x]}} + \sum_{\{i\;s.t\; \bar{c}\in C_i\}}{r_i[x]}
\end{equation}
Now, we analyze the efficiency of the method assuming that different parts of the noise are dominant.
\subsection{Poisson noise is dominant}
Assuming Poisson noise is the dominant noise source, we can rewrite equation \ref{eq:SNR} as $$\frac{MK}{N}F_{\bar{c}} \approx \sum_{\{i\;s.t\; \bar{c}\in C_i\}}{P(T^*_e g_{\bar{c}})}= P(\frac{MK}{N}T^*_eg_{\bar{c}})$$ 
We can choose $T^*_e = \frac{NT_e}{MK}$, so the equation becomes 
$$\frac{MK}{N}F_{\bar{c}} \approx P(\frac{MK}{N}T^*_eg_{\bar{c}}) = P(T_eg_{\bar{c}})$$
and we get the same SNR as in the original observation.
This means that the efficiency is $$E = \frac{N(T_e+T_r+T_s)}{M(T^*_e+T_r)+T_s} \approx \frac{NT_e}{MT^*_e} = \frac{NT_e}{M\frac{NT_e}{MK}}=K $$
This means that in both observing modes the multiplexed imaging gains maximal efficiency when the dominant noise is Poisson noise, and since there is no dependence on $M$ one can use large numbers of sub-observations guaranteeing high stability for the recovery algorithm.

\subsection{Background noise is dominant}\label{subsec:background}
Assuming background noise is the dominant noise source and the fact that $b_{i,j}$ are all independent, we can rewrite equation \ref{eq:SNR} above to 
$$\frac{MK}{N}F_{\bar{c}} \approx \frac{MK}{N}T^*_e g_{\bar{c}} + \sum_{\{i\;s.t\; \bar{c}\in C_i\}}{\sum_{j\in C_i}{\sqrt{T^*_e}b_{i,j}[x]}} =\frac{MK}{N}T^*_e g_{\bar{c}} + \sqrt{\frac{MK}{N}KT^*_e}b[x]$$
Now we can calculate the SNR when using multiplexed imaging:
$$SNR^* = \frac {\frac{MK}{N}T^*_e g_{\bar{c}}} {\sqrt{\frac{MK}{N}KT^*_e}\sigma_b} =\sqrt{\frac{M}{N}T^*_e}\frac{g_{\bar{c}}}{\sigma_b}  $$
Recall that the original SNR was $$SNR = \frac{T_eg_{\bar{c}}}{\sqrt{T_e}\sigma_b}=\sqrt{T_e}\frac{g_{\bar{c}}}{\sigma_b}$$
So, requiring equal SNR's, we need $T^*_e = \frac{NT_e}{M}$.
This means that the efficiency is 
$$E = \frac{N(T_e+T_r+T_s)}{M(T^*_e+T_r)+T_s}\approx \frac{NT_e}{MT^*_e} = \frac{NT_e}{M\frac{NT_e}{M}}=1 $$

It seems that the method does not help when the dominant noise is the background because the efficiency is $1$.
But for some applications (for example when performing shallow surveys) our assumption that $T_e>>T_r+T_s$ cannot be satisfied because the required exposure time for the observation is small compared to slew time or readout time. In these cases multiplexed imaging allows using $M$ exposures with a factor of $\frac{N}{M}$ larger exposure time instead of $N$ different exposures, with efficiency
$$E = \frac{N(T_e+T_r+T_s)}{M(T^*_e+T_r)+T_s}=\frac{N(T_e+T_r+T_s)}{M(\frac{N}{M}T_e+T_r)+T_s} = \frac{N(T_e+T_r+T_s)}{NT_e+MT_r+T_s} $$
which would mean $E=N$ if $T_s$ is dominant and $E=\frac{N}{M}$ if $T_r$ is dominant.

\subsection{Read noise is dominant}
When the read noise is dominant, equation \ref{eq:SNR} gives
$$\frac{MK}{N}F_{\bar{c}} \approx \frac{MK}{N}T^*_e g_{\bar{c}} + \sum_{\{i\;s.t\; \bar{c}\in C_i\}}{r_i[x]}  $$
The SNR when using multiplexed imaging is then:
$$SNR^* = \frac{\frac{MK}{N}T^*_e g_{\bar{c}}}{\sqrt{\frac{MK}{N}}\sigma_r} $$
So if we choose $T^*_e = \frac{T_e}{\sqrt{\frac{MK}{N}}}$ we get 
$$SNR^* = \frac{\frac{MK}{N}T^*_e g_{\bar{c}}}{\sqrt{\frac{MK}{N}}\sigma_r} = \frac{T_eg_{\bar{c}}}{\sigma_r} $$ which is equal to the original SNR.
From this we get the efficiency $$E = \frac{N(T_e+T_r+T_s)}{M(T^*_e+T_r)+T_s} = \frac{NT_e}{MT^*_e} = \frac{N\sqrt{\frac{MK}{N}}T_e}{MT_e}=\sqrt{\frac{NK}{M}}$$

In this case, the efficiency depends on the choice of $N,M$ and therefore there will be a difference between the charting mode and the re-observing mode. In re-observing, one needs only to extract the flux of most sources, neglecting overlapping stars and  therefore typically will use $N=K$ and $M=1$, meaning $E=\sqrt{\frac{KK}{1}}=K$. 
When charting, the efficiency depends on the parameters $N$, $M$ which are somewhat free to the choice of the system designer.

Note that to avoid having two indices $j_1,j_2$ such that $v_{j_1}=v_{j_2}$  we must have $\dbinom{M}{\frac{MK}{N}}\geq N$ (counting argument). This reproduces the obvious inequality on the efficiency $E=\sqrt{\frac{NK}{M}}\leq K$ (efficiency cannot be greater than $K$ as we observe at most $K$ areas at a time).
Note also that this boundary can be reached when $K << N$.
Using some reasonable parameters from our simulations, $M=2log_2(N),N=2K$ we get the efficiency $E= \frac{K}{\sqrt{log_2(N)}}$ which we expect will be typical for most uses.

\section{Simulations}\label{sec:Simulations}
\begin{figure}[H]
	\plottwo{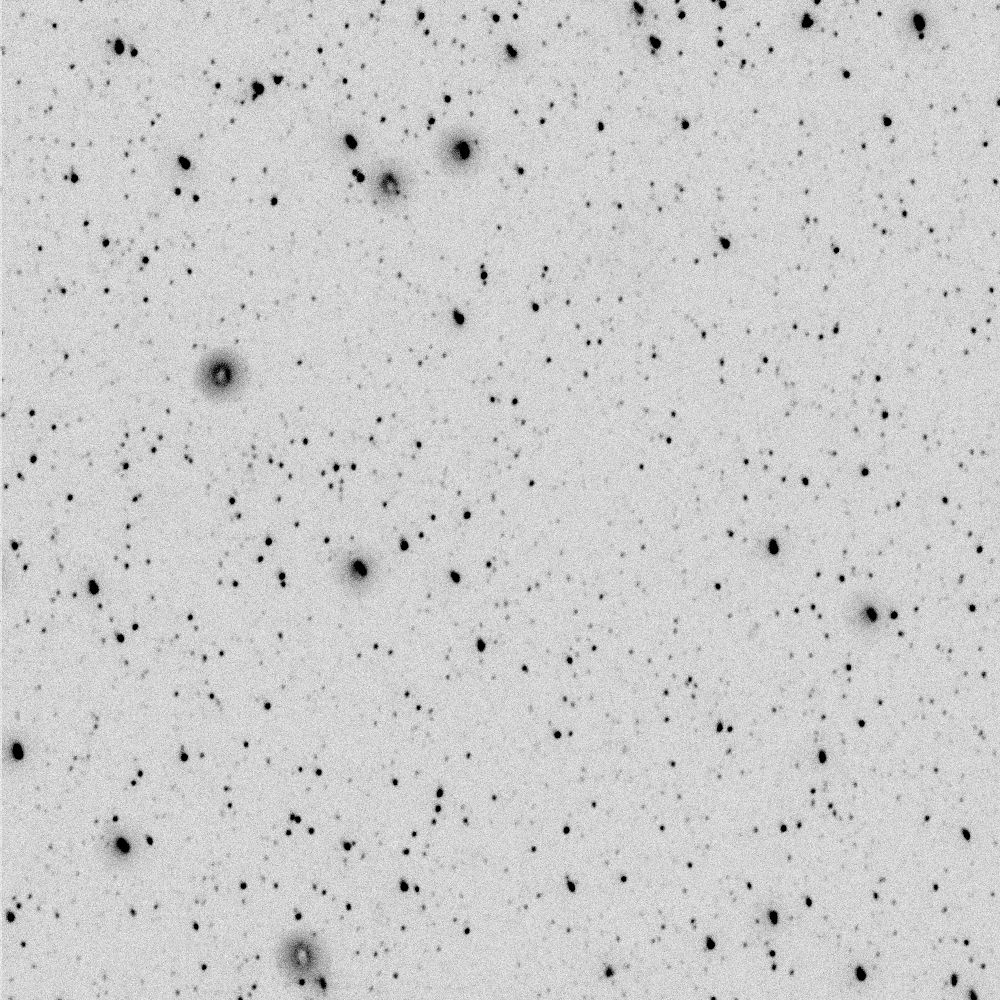}{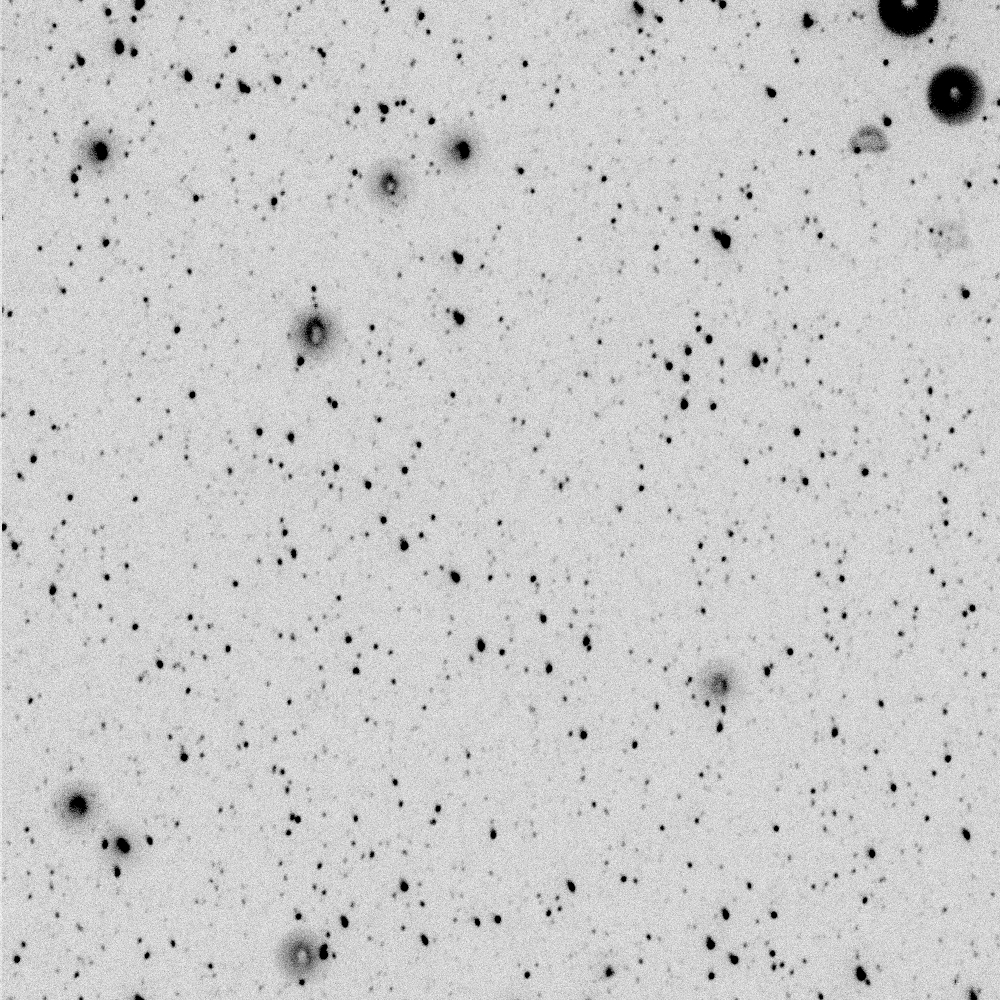}
	\caption{Examples of sub-observations using $N = 349$, $K\approx175$ and $M=18$.
  Notice that some objects appear in both images and some appear in only one of them.}\label{simulations_observations}
\end{figure}

To simulate the operation of the charting algorithm, we performed several simulations using data observed with the GALEX satellite (scanning mode with 80 seconds exposure time).
These observations were targeted at high galactic latitude, and therefore were sparse, with a measured $Pd\approx \frac{1}{1000}$ allowing for high multiplexing.
We used 349 regions, 1000x1000 pixels each.
We used our algorithm with $K=175$ and $M=18$ (see Fig. \ref{simulations_observations}). 
To study the performance of the recovery, we compare the reconstructed image with the original image with added background noise. Each reconstructed sky region is the average of 9 sub-observations in which it is contained. Therefore the theoretical standard deviation of the reconstructed image is $\sigma_{rec} = \frac{\sigma_{sub}}{\sqrt{9}}$. 
We generated our set $V$ satisfying conditions \eqref{couplesCondition} and \eqref{distanceCondition} with $r=4$ (the closest pair of vectors is different in 4 coordinates).

\begin{figure}[H]
\centering
\includegraphics[width=50mm]{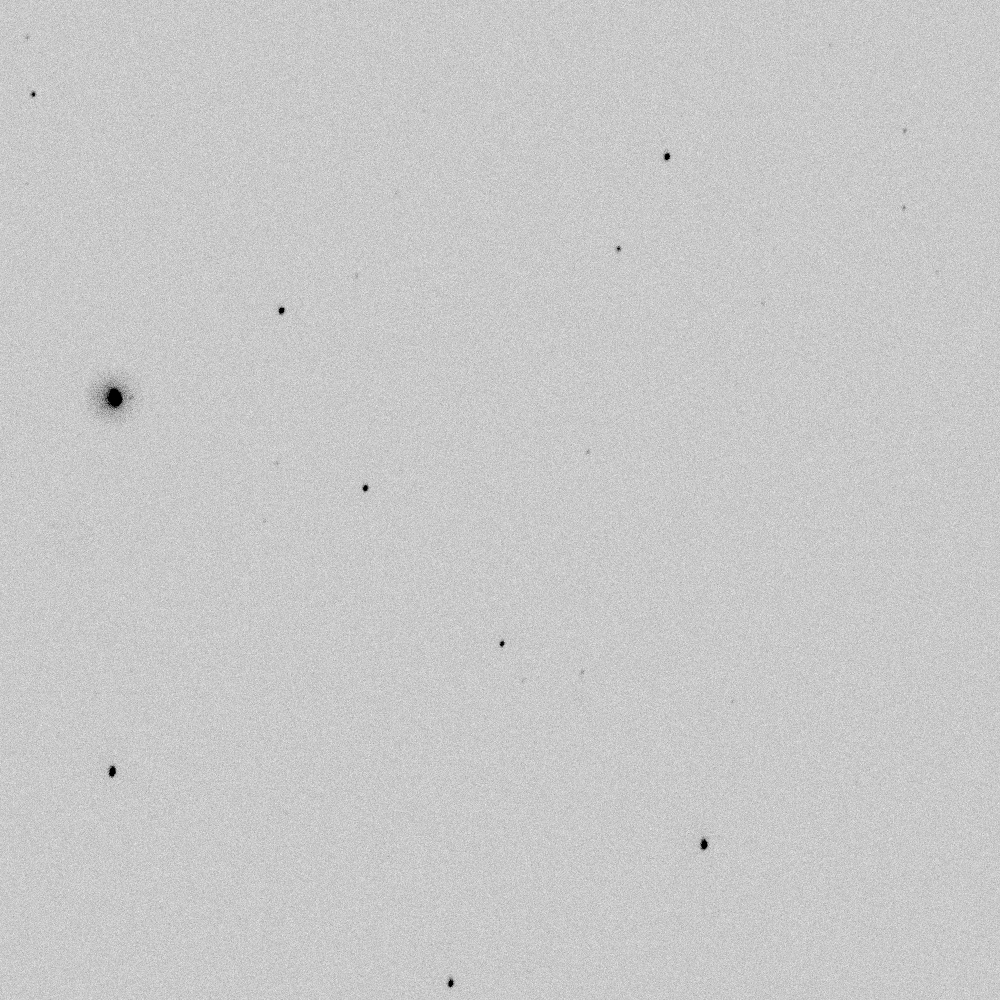}
\includegraphics[width=50mm]{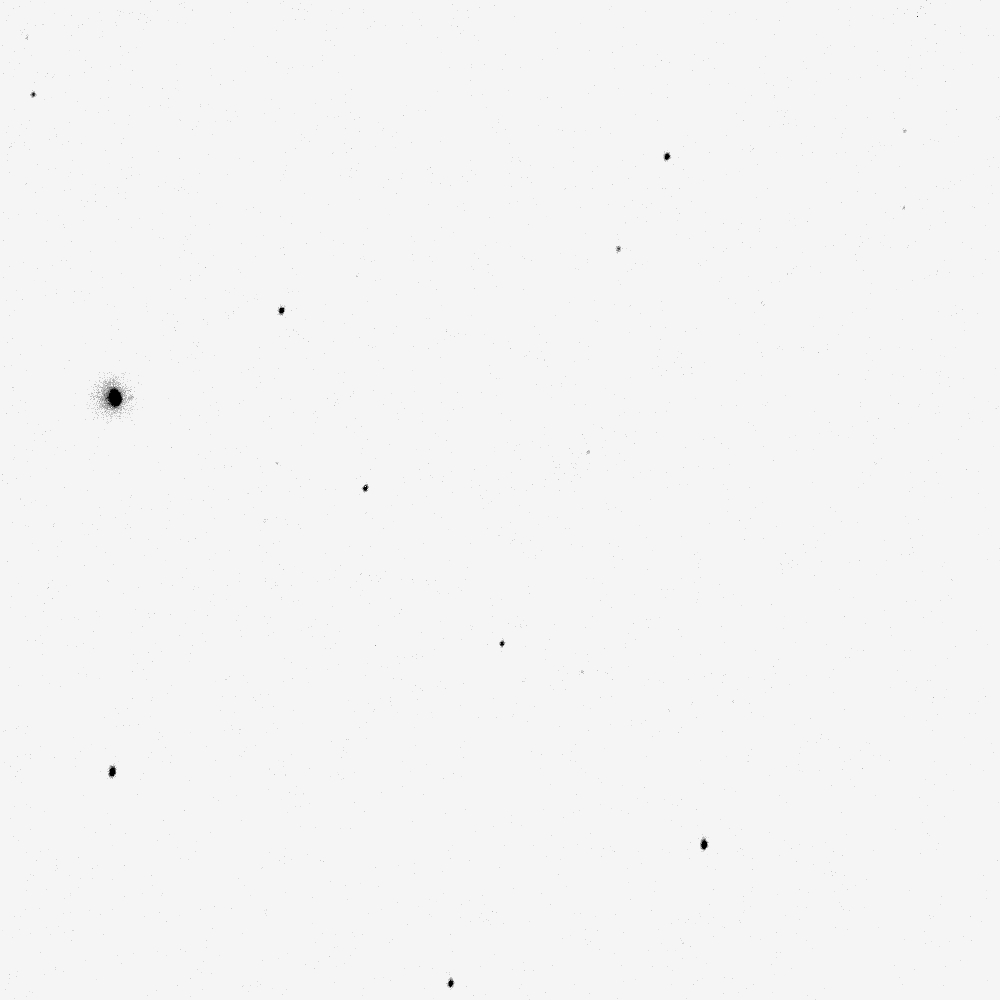}
\includegraphics[width=57mm]{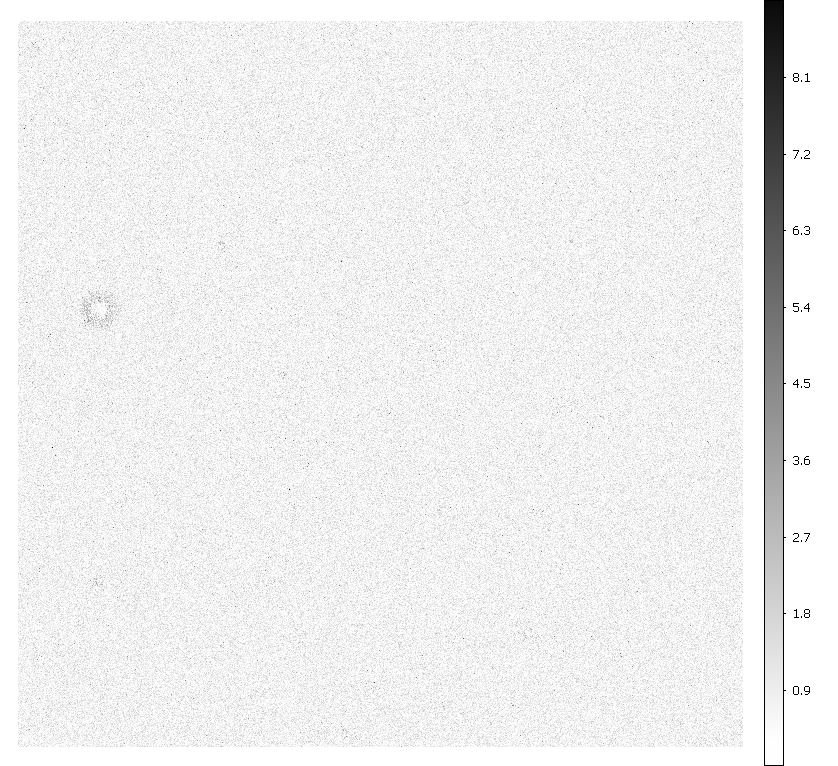}
\caption{Results of the charting recovery algorithm.
Upper left: The original image with added background noise to match the theoretical SNR of the reconstructed image. Upper right: The reconstructed image with the same gray-scale as the original image. Notice that only significant pixels have non-zero value. Bottom: The difference between them. The color-bar relates to the bottom image only and is in units of standard deviations of the background in the original image (upper left).}\label{fig:recovered_images}
\end{figure}

We report a high fidelity ($>95 \%$) for recovering the correct positions of all the $7\sigma_{rec}$  pixels which have flux contribution from 1 source (and $>99\%$ at $8\sigma_{rec}$ and above, see Fig. \ref{fig:recovered_images}).
We also report a high fidelity of recovering the correct combination of pixels with flux contribution of 2 or more sources ($>90\%$ for sources with more than $10\sigma_{rec}$ and $\sim99\%$ for sources with more than $20\sigma_{rec}$).

\section{Applications}\label{sec:Applications}
There are many possible applications to the multiplexed imaging method, and many scientific missions operating in the visible, UV and IR (from space) can substantially increase their capabilities by using it.
High multiplexing may be limited by the observed object density, or by practical/optical reasons. Here we discuss the expected improvements assuming no optical limitations, which will surely arise when designing a multiplexed imaging system with a large number of optical elements.
We assume that the object density ($Pd$) can be as low as $\frac{1}{1000}$, and the expected improvement shown below assume this object density.

\subsection {Sky surveys from space}\label{subsec:Sky surveys from space}
Multiplexed imaging would be powerful for sky surveys from space because of the combination of the following factors:
\begin{itemize}
\item Detectors are more expensive to operate in space, and the multiplexed imaging system may reduce the amount of expensive space-qualified hardware.
\item The background noise is substantially lower from space than from the ground. 
\item There are no atmospheric aberrations, meaning that the PSF when imaging from space is substantially smaller, reducing $Pd$, allowing for higher multiplexing.
\end{itemize}
\textbf {Expected improvement:} Our simulations suggest one can use multiplexing as high as $K=175$ with charting mode (section \ref{sec:Simulations}), leading to increased area coverage per unit time by a factor of $\sqrt{\frac{350*175}{18}}\approx 60$. 

\subsection {Lucky imaging}\label{subsec:Lucky imaging}
The lucky imaging method is a technique used to decrease the effects of atmospheric aberrations using high frequency imaging \citep{luckyImaging}.
When imaging at high frequency, the dominant noise source is the read noise, meaning that high multiplexing might be useful.
One should notice that the current charting recovery algorithm will not suffice for this approach, because objects that fall on the same detector area will interfere differently on each sub-observation, as a result of the atmospheric aberrations. 
Therefore, we must assume that the logical limit set by the object density is lower than the bound we got from the object density, to prevent the interference of objects.
Since high-speed detectors are small and expensive, using the multiplexed imaging method the lucky imaging technique may become more useful for imaging larger sky areas.\newline
\textbf{Expected improvement:} An increase in the range of 100 fold (assuming we cannot allow colliding sources) to 1000-fold (assuming we can) in the area observed.

\subsection {Fast photometry}\label{subsec:Fast photometry}
Astronomers use high frequency observations ($\sim 50Hz$) to search for rapid changes in the light flux coming from stars, e.g caused by random occultations by Kuiper belt objects \citep{kuiper} or by intrinsic changes of the stellar flux (e.g, astro-seismology, \citealp{astroseismology}).
In this scientific use, the dominant source of noise is either the read-noise or the Poisson noise, allowing for high multiplexing.\newline
\textbf{Expected improvement:} An increase of up to 1000 in the amount of stars we will be able to monitor.

\subsection {Searching for transient sources}\label{subsec:Searching for transients}
When searching for transients, one seeks the appearance of new objects in the field of view. This means that in principle astronomers try to image as wide an area as possible, observing the same area of sky over and over again. With multiplexed imaging, one can use the re-observing mode.\newline
\textbf{Expected improvement from the ground:} From the ground, when the background noise is dominant, multiplexing can help making shallow all-sky surveys 10-fold to 100-fold more effective, increasing the cadence and reducing the efficiency drop due to slew time. 
When designing new instruments, multiplexed imaging may help reduce the cost of survey telescopes, allowing for the use of smaller detectors, larger f-numbers and reducing the demands from the physical machinery.\newline
\textbf{Expected improvement from space:} In this case the dominant noise source is either read-noise or Poisson noise. The expected improvement is up to a factor of 1000.

\subsection {Searching for planets, eclipsing binaries and micro-lensing events}\label{subsec:Planet search}
These applications involve monitoring bright stars regularly, to detect flux decrements due to occultation of the star by a planet or an increase of flux due to a lensing event. The flux variability scale can be as small as 0.0001\%. At these levels of precision the dominant noise is the Poisson noise, allowing for high multiplexing. It will be especially beneficial when making shallow homogeneous searches for planets.\newline
\textbf{Expected Improvement:} Improvement factor of up to 1000 when searching non-dense sky areas, and roughly 10 when monitoring dense regions of the sky.

\section{Summary}\label{sec:summary}
Multiplexed imaging systems could be game-changers for future space missions, and would be useful also for new ground based instrumentation.

The use of the method may lead to a new generation of wide-field surveying space telescopes as well as efficient ground-based instruments for lucky imaging, fast photometry, and transient and variability surveys. 

\section{Supplementary}
\subsection{Construction of the sets}\label{subsec:construction}
In this section we show how one can construct the set $V$ such that  there will be no ambiguity in the recovery of two interfering sources.
We will start by noticing a useful fact:

{\bf Claim.}
{\it Let $\overrightarrow{\mu}$ be a linear combination of two binary vectors, $\overrightarrow{\mu} = c\overrightarrow{v_0}+d\overrightarrow{w_0}$, such that all the numbers in the set $\{a,b,a+b\}$ appear in $\overrightarrow{\mu}$. Then the only set of numbers $\{c,d\}$ such that $\exists \ora{v},\ora{w}$ binary vectors and $\overrightarrow{\mu} = c\overrightarrow{v}+d\overrightarrow{w}$ is $\{a,b\}$.}

{\bf Proof.} Assume that there exists $c,d,\ora{v},\ora{w}$ such that $\mu = c\ora{v} + d\ora{w}$.
W.l.o.g, $a<b$ and $c<d$.
Since $\ora{v},\ora{w}$ are binary vectors, the only numbers that can appear in $\ora{\mu}$ are $0,c,d,c+d$. So the sets $\{0,a,b,a+b\},\{0,c,d,c+d\}$ must be equal, but since both sets are sorted, this means that 
 $a=c,b=d$.
{\newline \it Q.E.D \newline}

If we assume that $\frac{|a-b|}{\sqrt{a+\sigma_{noise}^2}}>\gamma$, we know that the numbers $\{0,a,b,a+b\}$ are statistically distinguishable in our measurement, meaning that for each index $i$ in the vector $\mu$ we can decide which of the values $\{0,a,b,a+b\}$ $\mu[i]$ gets.
This means that we can exactly construct $\mu$ as $\mu = av_j + bv_l$ with no confusion, leading to a correct recovery of the indices $j,l$.
In order to construct the set $V$ we will introduce another helpful fact.

{ \bf Claim.} {For every pair of binary vectors $v\neq w$ with equal sum $K$ where $K>M/2$, all the numbers $a,b,a+b$ appear in the sum $a\overrightarrow{v}+b\overrightarrow{w}$.}

{\bf Proof.} The sum of $v+w$ is $2K$ which is distributed over $M$ coordinates, and $2K>M$ meaning $v+w$ contains a coordinate with weight larger than 1, so  $a\overrightarrow{v}+b\overrightarrow{w}$ contains the number $a+b$. Since $v\neq w$ there is at least one coordinate where they are different and they both have the same number of 1's, so they must be different on an even number of places, therefore $a$ and $b$ also appear in $a\overrightarrow{v}+b\overrightarrow{w}$.
{\newline \it Q.E.D \newline}

To assure that all the numbers $\{a,b,a+b\}$ appear in every combination of $av_j+bv_l$ for every $v_j\neq v_l \in V$ we can choose the set $V$ to be a set of binary vectors with equal sum, and which will be chosen to be $\frac{MK}{N}$.
We need to have $|V| = N$, therefore we can determine $M$ by the relation $N< \binom{M}{\frac{MK}{N}}$, deriving that $$M = \alpha log(N) + \beta log(log(N))$$ 
Using an example ratio of $\frac{K}{N} = \frac{1}{2}$, we  get $\alpha = 1, \beta = 0.5$.

If we want to recover the position in the situation where $a=b$ (or $a$ and $b$ are indistinguishable due to noise) there needs to be only one way to recover $v,w\in V$ from $v+w$. We may also want to enforce that every pair of vectors are sufficiently distant to help reduce the confusion in the positions of weak sources.
Both conditions can be satisfied by constructing the set $V$ from the empty set by inserting vectors with exactly $\frac{MK}{N}$ ones in a random order, verifying the condition $v_1+v_2 \neq w_1 + w_2$ for every quadruplet of $V$ and the condition $||v_i-v_j||\geq r$ at every time.
Experimentally we got that the achieved $M$ with this process is roughly $M \approx 2log(N)$ with $r=4$, which is about twice the value of $M$ needed without the above limiting conditions.

\subsection{Reconstruction algorithm}\label{Algorithm} 
Define the confidence parameter $\gamma\approx 5$.
For every pixel $x$ on the detector: 
\begin{enumerate}
  \item Let $\ora{\mu}$ be the vector of all its sub-observations. Initialize the set $W = \emptyset$.
  \item Construct the error vector $\overrightarrow{\sigma_x}[i] = \sqrt{\overrightarrow{\mu}[i] + K\sigma_{background}^2 + \sigma_{read}^2 }$ and set the loss $S = ||\frac{\mu}{\sigma_x}||^2$ .

  \item For every $v\in V$ use weighted least squares to find the parameters $\alpha_i,\beta$ such that the loss $\abs{\frac{\overrightarrow{\mu} -\sum_{\ora{w_i}\in W}{\alpha_i\ora{w_i}} - \beta\overrightarrow{v}}{\overrightarrow{\sigma_x}}}^2$ is minimal. Choose the pair $v,\beta$ that minimizes the loss.
  \item If $S-\abs{\frac{\overrightarrow{\mu} -\sum_{\ora{w_i}\in W}{\alpha_i\ora{w_i}} - \beta\overrightarrow{v}}{\overrightarrow{\sigma_x}}}^2 > \gamma^2$
  \begin{enumerate}
  \item Add v to the chosen vectors set $W$.
  \item Set the loss $S = \abs{\frac{\overrightarrow{\mu} -\sum_{\ora{w_i}\in W}{\alpha_i\ora{w_i}} - \beta\overrightarrow{v}}{\overrightarrow{\sigma_x}}}^2$  
  \item Repeat stage 3.
  \end{enumerate}
  \item Use the weighted least squares algorithm to find $\alpha_i$  that minimizes $\abs{\frac{\overrightarrow{\mu} -\sum_{\ora{w_i}\in W}{\alpha_i\ora{w_i}}}{\overrightarrow{\sigma_x}}}^2$, and output all the couples $\ora{w_i},\alpha_i$.

\end{enumerate}

\end{document}